\documentclass[12pt]{article}
\usepackage{color,graphicx,amssymb,amsmath}

\newcommand{\makebf}[1]{{\mbox{\bf\boldmath ${#1}$}}}

\newcommand{\bfgm}{{\makebf{\gamma}}}

\newtheorem{remark}{Remark}
\newtheorem{lemma}{Lemma}
\newtheorem{example}{Example}
\newtheorem{theorem}{Theorem}

\title{An order-preserving property of additive invariant for Takesue-type reversible cellular automata}
\author{Gianluca Caterina\footnote{Present address:  Department of Mathematics, Northeastern University, 360 Huntington Avenue, Boston, MA 02115, USA}\; and Bruce M. Boghosian\\
{\scriptsize\it Department of Mathematics, Tufts University, 503 Boston Avenue, Medford, Massachusetts 02420, USA}
}
\begin{document}
\maketitle
\abstract{We show that, for a fairly large class of reversible, one-dimensional cellular automata, the set of additive invariants exhibits an algebraic structure. More precisely, if $f$ and $g$ are one-dimensional, reversible cellular automata of the kind considered by Takesue \cite{Takesue}, we show that there is a binary operation on these automata $\vee$ such that $\psi(f)\subseteq \psi(f\vee g)$, where $\psi(f)$ denotes the set of additive invariants of $f$ and $\subseteq$ denotes the inclusion relation between real subspaces.}


\section{Conserved quantities and symmetries}
    
Additive invariants for cellular automata (CA) have been widely studied over the last two decades, especially in connection with the fundamental role they play in physical modeling. The problem of finding explicit conditions for a one-dimensional CA to have conserved densities of a given size (in a sense that will be described in this article) has been solved by Hattori and Takesue in their remarkable 1991 paper~\cite{hattak}. Interestingly, they expressed those conditions in terms of discrete current laws, thereby recovering a fundamental property of continuous systems in a discrete context. 

This striking analogy with continuous models suggested that a discrete version of Noether's theorem -- an elegant and fundamental connection between conserved quantities and symmetries of physical systems --  might be formulable.  On the other hand, the proof of Noether's theorem is based on the properties of differential operators which are absent in a discrete context.

For these reasons we decided to approach this problem first from a computational point of view. We considered a class  $\mathcal{F}_I$ of second-order, reversible CA based on a binary alphabet  defined over a fixed ``neighbor window'' $I$ and, for each $f\in \mathcal{F}_I$, computed all its additive conserved quantities, the set of which may be identified with a finite-dimensional vector space denoted by $\psi(f)$. Then we looked at the equivalence classes of  CA rules given by the relation

\[f \sim g \iff \psi(f)=\psi(g)\]
and observed that, if we choose two representatives $f$ and $g$ from the same equivalence class $\psi(f)$, then there is a pointwise binary operation $\vee$ on $\mathcal{F}_I$ such that  
\begin{equation}\label{eq: 23}
\psi(f)\subseteq\psi(f\vee g).
\end{equation}
This relation is interesting because it relates a dynamical system to its invariants. Since Noether's theorem endeavors to relate the latter with the symmetries of the system, we thought it was worth investigating this problem further. In this work we present a proof of Eq.(\ref{eq: 23}) for a large class of reversible, second-order dynamical systems defined on an arbitrary finite set.

\section{Physical motivations}

In a seminal paper~\cite{Takesue}, Takesue argues that a family of fully-discrete models called Elementary Reversible Cellular Automata (ERCA) can be used to study the thermodynamic behavior of large dynamical systems. ERCA are a family of one-dimensional reversible CA with two Boolean independent variables at each site. It is possible to associate to these systems certain additive conserved quantities which can be regarded as a form of energy. Also, discreteness of the variables  implies that the phace space volume is preserved under the dynamics, and therefore the statistical mechanics of the model can be constructed. 

Examples of physical applications of these quantities are given, for instance, by Pomeau~\cite{Pomeau} and Boccara~\cite{boccara, boccara1} who have used them successfully to model traffic. In most of these works, special symmetries of the rules are considered and conjectures are drawn on how they relate to the associated set of additive invariants. 

In our approach, we give primacy to the equivalence relation which identifies two rules if they have the same set of invariants. The main result is a theorem which shows that symmetries of the rules are reflected in an algebraic structure on their respective equivalence classes. Since ``energies'' are hydrodynamic quantities which one hopes to see emerging from the dynamics, our result can be seen as a contribution toward understanding how certain qualities of the dynamics reflect on the symmetries of the associated physically relevant quantities. In the next section we start the discussion by generalizing ERCA to an arbitrary, finite alphabet.

\section{Generalized second-order, reversible CA}

The instantaneous state of many discrete dynamical systems is most conveniently described by $n$ dependent variables, each of which takes its values in a set $S$, with $|S|=N< \infty$.  
We denote the $j$'th dependent variable at time step $t$ by $x_j(t)\in S$, for $0\leq j<n$ and $t\geq 0$, so that the state of the system at time $t$ 
\[
x(t)=\left\{x_0(t),x_1(t),\dots, x_{n-1}(t)\right\}
\]
may be thought of as an $n$-vector. CA dynamical systems of this kind endowed with a shift-invariant map \cite{hed}. 

More precisely, consider the set $I_{\beta}=\{-\beta, -\beta+1,\dots,0,\dots,\beta -1, \beta\}$ ($\beta\in \mathbb{Z}$) and a function
 \[f: S^{I_\beta} \longrightarrow S.\]
 Then a one-dimensional cellular automaton is the dynamical system defined by the triple 
 \[(S,\beta,f),\]
 with the dynamics being described by the following map:
\[
x_{i}(t+1)=f(x_{i-\beta}(t),\dots,  x_i(t), \dots, x_{i+\beta}(t)).
\]
By defining $\hat{x}_i^{\beta}(t):=\{x_{i-\beta}(t),\dots, x_{i}(t),\dots,x_{i+\beta}(t)\}$, the above can be rewritten as
\[
x_i(t+1)=f(\hat{x}_i^{\beta}(t)).
\]
We refer to the associate global map $F:x(t)\mapsto x(t+1)$ as the global map induced by $f$.

In what follows, we will also assume {\it periodic boundary conditions}, which means that $x_{(i+j)}=x_{(i+j) mod\ n }$ and we refer to the system so defined as a cellular automaton (CA) of {\it first order} and neighborhood size  $\beta$.

We now define the concept of second-order reversible CA, mentioned in the Introduction. Let 
\[\gamma: S^2\longrightarrow S\]
be a second-order reversible map; that is $\gamma$ satisfies
 \[
 \forall_{x,y\in S}:\gamma(y,\gamma(y,x))=x.
 \] 
Then we can define
\[
\tilde{f}:S^{I_\beta}\times S \longrightarrow S
\]
\begin{eqnarray}\label{eq:secondorder1}(\hat{x}_{i}(t+1), x_i (t))\mapsto\gamma(f(\hat{x}_i(t+1)),x_i(t))=x_{i}(t+2).\end{eqnarray}
Consistent with the notation introduced above, we call $\tilde{f}$ a {\it second-order} CA of neighborhood size $\beta$ and {\it reversible combiner} $\gamma$. Since the binary version of these objects was first introduced in \cite{hattak}, we will also refer to them as {\it Takesue-type} cellular automata. When it will be clear from the context, we will drop the tilde from the notation. 

We notice that:
\begin{eqnarray}\nonumber\label{eq:secondorder2}\tilde{f}(x_{i}(t+1),x_i(t+2))=\gamma(f(\hat{x}_i(t+1)),x_i(t+2))=\\=\gamma(f(\hat{x}_i(t+1)),\gamma(f(\hat{x}_i(t+1)),x_i(t)))=x_i(t).\end{eqnarray}
Hence, by Eqs.(\ref{eq:secondorder1}) and (\ref{eq:secondorder2}), we deduce that the second-order dynamical system $\tilde{F}$ induced by $\tilde{f}$,
\[\tilde{F}(x(t+1),x(t))=x(t+2)\]
implies 
\[\tilde{F}(x(t+1),x(t+2))=x(t)\]
demonstrating that $\tilde{F}$ is indeed a reversible dynamical system. 

 \section{Additive conserved quantities} 

Consider a second-order reversible CA $f$ of neighborhood size $\beta$ on a periodic lattice of size  $N$. We define an ``energy density function''  
\[
\epsilon:S^{2\alpha+1}\times S \longrightarrow \mathbb{R}
\]
\[
(\hat{x}_i^{\alpha}(t),x_i(t-1))\mapsto\epsilon(\hat{x}_i^{\alpha}(t),x_i(t-1)).
\]
\begin{remark}
In what follows, two distinct neighborhoods will appear: One is relative to the automaton (neighborhood size $\beta$), and one is relative to the conserved density (neighborhood size $\alpha$). However, the $\alpha$-neighborhood will appear exclusively as an argument of $\epsilon$, whereas the $\beta$-neighborhood will appear only as a property of the automaton. Hence, in order to simplify the notation, we can drop the superscripts $\alpha$ and $\beta$ with no ambiguity.
\end{remark}
We define the {\it total energy} $E_N$ of the system at time $t$ associated to $\epsilon$ as:
\begin{equation}
E_N(x(t),x(t-1),\epsilon)=\sum_{i=0}^{N-1}\epsilon(\hat{x}_i(t),x_i(t-1)).
\end{equation}
When the density $\epsilon$ is given, or clear from the context, it will be dropped from the arguments for $E$. 

We say that $\epsilon$ is a {\it additive conserved density for $\tilde{f}$} if, for any $t>0$ and $N>0$ we have 
\begin{equation}\label{eq:conditionsrevers}
E_N(x(t),x(t-1),\epsilon)=E_N(\tilde{F}(x(t),x(t-1)),x(t),\epsilon).
\end{equation}
According to this definition, in order for $\epsilon$ to be a conserved quantity, an infinite number of conditions must hold: namely, conserved quantities are solutions to the system of equations defined by Eq.~($\ref{eq:conditionsrevers})$ for any $N\in\mathbb{N}^+$. In their seminal paper~\cite{hattak} Hattori and Takesue proved that only a finite number of linear conditions are sufficient to find all the solutions to Eq.~$(\ref{eq:conditionsrevers})$. 

These conditions can be rephrased in a slightly different, though equivalent, fashion. Indeed, let us denote by $\psi(\tilde{f},N,\alpha)$ the set of all the additive conserved densities of neighborhood size $\alpha$ for $\tilde{f}$ on a periodic lattice of size $N$. Then we have:
\begin{lemma}\label{le:le1}
If there is $N_0$ such that $\psi(\tilde{f},N_0,\alpha)=\psi(\tilde{f},N_0+1,\alpha)$ then $$\psi(\tilde{f},M,\alpha)=(\tilde{f},N_0,\alpha),\  \forall M>N_0$$
\end{lemma}
\begin{lemma}\label{le:le2}
For any $\tilde{f}$ and any $\alpha$ there exists $N_0\leq\infty$ such that $$\psi(\tilde{f},N_0,\alpha)=\psi(\tilde{f},N_0+1,\alpha).$$
\end{lemma}
For the proof of these lemmas we refer to~\cite{hattak}. Let us notice that the above result tells us that the set of invariants for a given automaton $\tilde{f}$ is, for $N$ sufficiently large,  independent of the size of the periodic lattice on which the system evolves. Therefore we can drop the dependence on $N$ and define the set of additive conserved quantities for $\tilde{f}$ to be $\psi(\tilde{f},\alpha)=\psi(\tilde{f},N_0,\alpha)$. 

\section{An algebraic property of additive invariants}

In this section we prove that additive invariants for the second-order, reversible CA considered above possess a natural equivalence class structure.
\begin{remark}
The choice for the {\it space-time} window neighborhood  (in particular, the choice of the ``central'' point of the neighborhood at time $t-2$) which defines the density $\epsilon$ is crucial to our analysis. The idea is to force the window for $\epsilon$ to have the same ``shape''  as the one for the second-order automaton $\tilde{f}$: In the next section the consequences of this choice on our argument will appear more clearly.
\end{remark}

Using the notation introduced above, let us consider a second-order reversible CA with $S=\{0,1\}$, 
$\beta=1$ and $\alpha=1$. This corresponds to the following time evolution:
\[x_i(t+2)=f(\hat{x}_i(t+1))\oplus x_i(t)\]
where $x_i(t)\in\{0,1\}$ for all $i,s,$ and $\oplus$ is addition modulo $2$. Since we set $\alpha=1$, the density $\epsilon$ has the form $\epsilon=\epsilon(x_{i-1}(t), x_i(t), x_{i+1}(t), x_i(t-1))$ and can therefore be represented by a vector with 16 components. 

Let us now fix the window: That is, let us denote by $\mathcal{F}$ the set of all the functions
\[\{0,1\}^3\longrightarrow \{0,1\},\]
and, for any $f\in\mathcal{F}$, let $\phi(f)$ be the set of all second-order additive conserved densities for $\tilde{f}$ with $\alpha=1$.  The equivalence relation
\[f\sim g\iff \psi(f)=\psi(g)\]
then induces a partition of $\mathcal{F}$ into equivalence classes. We did an exhaustive computer search and found that $\sim$ partitions $\mathcal{F}$ into 21 distinct equivalence classes. 
\begin{example}
The following independent row vectors, written in matrix form as:
$$\left(\begin{array}{cccccccccccccccc} 
1&0&0&0&2&1&2&2&0&0&-1&0&0&0&0&1 \\
 0 &1&1&1&0&1&1&1&0&0&1&0&0&0&1&0 \\
-1&0&0&0&-2&0&-2&-1&0&0&1&0&0&1&0&0 \\
1&0&0&0&2&0&1&0&0&0&-1&0&1&0&0&0 \\
0&0&1&1&-1&-1&-1&-1&0&0&1&1&0&0&0&0 \\
0&0&-1&-1&0&0&0&0&1&1&0&0&0&0&0&0 \\
\end{array}\right)$$
are all the possible conserved quantities of  the CA represented, using Wolfram's notation \cite{Wolfram1}, by 8,64 and 72. Therefore $\{8,64, 72\}$ comprises a class in $\mathcal{F}/\sim$.
\end{example}

In the above example, CA 8, 64 and 72 correspond, respectively, to the following binary sequences:
\[8\mapsto (0,0,0,0,1,0,0,0),\ 64 \mapsto (0,0,0,1,0,0,0,0),\ 72\mapsto (0,0,0,1,1,0,0,0)\]
If we define $f\vee g$ to be the bitwise inclusive $or$ of the binary representations of $f$ and $g$, we can easily verify that the class formed by $8,64,72$ is closed with respect to this operation.  
Interestingly, this closure property holds for all the classes except two. However, the following slightly weaker property holds in this context:
\newtheorem{conjecture}{Conjecture}\label{conj:conj1}
\begin{conjecture}
If $f,g$ belongs to the same equivalence class, then we have that 
\[\psi(f)\subseteq\psi(f\vee g).\]
\end{conjecture}
This is interesting, since it says that there is an ``invariant-preserving'' operation defined on the class of CA with a given window function. In the next section we will present a lemma which will be used to prove, in Section~\ref{sec:theorem},  a generalization of this result to an arbitrary finite alphabet.

\section{The main lemma}

Following the notation introduced in the previous section, let $\tilde{f}$ be a second order reversible cellular automaton defined by 
\[\tilde{f}:S^{2\beta+1}\times S\longrightarrow S\]
\[(x_i(t),x_i(t-1))\mapsto\gamma(f(x_i(t)),x_i(t-1)),\]
where
\[\gamma:S^2\longrightarrow S\]
is such that
\[
\forall_{x,y\in S}:\gamma(y,\gamma(y,x))=x.
\]
Notice that, on a binary alphabet, the only non-trivial function with this property is addition modulo 2 (up to conjugation), which we denote by $\oplus$. Notice also that, since $\oplus$ is associative and commutative, it satisfies 
\[
(x\oplus (y\oplus z))=(y\oplus (x\oplus z)).
\]
The next lemma shows that these two properties of the reversible combiner are sufficient to prove Conjecture~\ref{conj:conj1}.
\begin{lemma}
\label{prop:prop1}
Suppose that  $\tilde{f}$ is a second-order reversible cellular automaton and
\begin{enumerate}
\item $\gamma(y,\gamma(y,x))=x$ 
\item $\gamma(x,\gamma(y,z))=\gamma(y,\gamma(x,z))$.
\end{enumerate}
Then, if $\epsilon$ is an additive conserved quantity density for $\tilde{f}$ and $\tilde{g}$ we have that
\begin{equation}
\epsilon[\hat{x}_i(t),\gamma(f(\hat{x}_i(t)),x_i(t-1))]= \epsilon[\hat{x}_i(t),\gamma(g(\hat{x}_i(t)),x_i(t-1))]
\end{equation}
for any $i\in \{0,1,\dots,N-1\}$, $t\geq 1.$
\end{lemma}
{\it Proof}. Since the system is reversible it is enough to prove the claim for case $t=1$.

First notice that, if $f(\hat{x}_i(1))= g(\hat{x}_i(1))$, the claim is trivial. Then, let $x(0)=\{x_0(0),x_1(0),\dots,x_{N-1}(0)\}$ and $x(1)=\{x_0(1),x_1(1),\dots,x_{N-1}(1)\}$ be the initial conditions. 

We set $x_f(2):=\tilde{F}(x(1),x(0))$ and define the change of energy forward in time 
\[\Delta_{N,f}(x(1),x(0))=E_N(x_f(2),x(1))- E_N(x(1),x(0)),\]
and the change of energy backward in time
\[\Delta'_{N,f}(x(1),x(0))=E_N(x(0),x(1))- E_N(x(1),x_f(2)).\]

Let $I=\{i_1,i_2,\dots,i_k\}$ be the set of all $i$ such that $f(\hat{x}_i(1))\neq g(\hat{x}_i(1))$ and fix an index $j\in I$.
Since $\epsilon$ is a conserved quantity for both $\tilde{f}$ and $\tilde{g}$, we have that
\[
\Delta'_{N,f}(x(1),x_f(2))-\Delta'_{N,g}(x(1),x_f(2))=0,
\]
and hence we have
\begin{eqnarray}
\label{eq:prima}
\lefteqn{\phantom{-}\left[
\sum_{i=0}^{N-1}\epsilon(\hat{x}_{i}(0),x_{i}(1))-\sum_{i=0}^{N-1}\epsilon(\hat{x}_i(1),\gamma(f(\hat{x}_i(1)),x_{i}(0)))\right]}\nonumber\\
\lefteqn{-\left[\sum_{i=0}^{N-1}\epsilon(\hat{x}_{i}(0),x_{i}(1))-\sum_{i=0}^{N-1}\epsilon(\hat{x}_i(1),\gamma(g(\hat{x}_i(1)),x_{i}(0))\right]}\nonumber\\
&=&
\left[\sum_{i\in I \backslash j}\epsilon(\hat{x}_i(1),\gamma(f(\hat{x}_i(1)),x_{i}(0))-\epsilon(\hat{x}_i(1),\gamma(g(\hat{x}_i(1)),x_{i}(0)))\right]\nonumber\\
& &
+\epsilon(\hat{x}_{j}(1),\gamma(f(\hat{x}_{j}(1)),x_j(0)))-\epsilon(\hat{x}_{j}(1),\gamma(g(\hat{x}_{j}(1)),x_j(0)))\nonumber\\
&=& 0.
\end{eqnarray}

Consider now $x(0)$ and replace $x_j(0)$ by $\gamma(f(\hat{x}_{j}(1)),\gamma(g(\hat{x}_{j}(1)),x_j(0)))$. We denote this new state by $\overline{x}(0)$ and consider the evolution with initial conditions $(x(1),\overline{x}(0))$ under the dynamics induced by $f$. At time $t+2$, the value of the $j^{th}$ dependent variable is equal to 
\begin{eqnarray}
\gamma(f(\hat{x}_{j}(1)),\gamma(f(\hat{x}_{j}(1)),\gamma(g(\hat{x}_{j}(1)),x_j(0))))=\gamma(g(\hat{x}_{j}(1)),x_j(0)).
\end{eqnarray}
Here we have used $\gamma(y,\gamma(y,x))=x$, with $y=f(\hat{x}_{j}(1))$ and $x=\gamma(g(\hat{x}_{j}(1)),x_j(0))$.

By a similar argument, if we consider the evolution with initial conditions $(x(1),\overline{x}(0))$ under the dynamics induced by $g$, at time $t+2$, the value of the $j^{th}$ dependent variable is equal to 
\begin{eqnarray}
\lefteqn{
\gamma(g(\hat{x}_{j}(1)),\gamma(f(\hat{x}_{j}(1)),\gamma(g(\hat{x}_{j}(1)),x_j(0))))}\phantom{aaa}\nonumber\\
&=&
\gamma(g(\hat{x}_{j}(1)),\gamma(g(\hat{x}_{j}(1)),\gamma(f(\hat{x}_{j}(1)),x_j(0))))\nonumber\\
&=&
\gamma(f(\hat{x}_{j}(1)),x_j(0)).
\end{eqnarray}
Here we have used the fact that, by hypothesis,  $\gamma(x,\gamma(y,z))=\gamma(y,\gamma(x,z))$ and therefore
\[\gamma(f(\hat{x}_{j}(1)),\gamma(g(\hat{x}_{j}(1)),x_j(0)))=\gamma(g(\hat{x}_{j}(1)),\gamma(f(\hat{x}_{j}(1)),x_j(0))).\]
Because of this, for initial conditions $(x(1),\overline{x}(0))$, conservation of energy gives rise to the identity
\[\Delta'_{N,f}(x(1),\overline{x}(0))-\Delta'_{N,g}(x(1),\overline{x}(0))=0,\]
or
\begin{eqnarray}\label{eq:seconda}
\lefteqn{\left[\sum_{i\in I \backslash j}\epsilon(\hat{x}_i(1),\gamma(f(\hat{x}_i(1)),x_{i}(0)))-\epsilon(\hat{x}_i(1),\gamma(g(\hat{x}_i(1)),x_{i}(0)))\right]}\phantom{aaaaaa}\nonumber\\
& &
+(\epsilon(\hat{x}_{j}(1),\gamma(g(\hat{x}_{j}(1)),x_j(0)))-\epsilon(\hat{x}_{j}(1),\gamma(f(\hat{x}_{j}(1)),x_j(0)))\nonumber\\
&=&
0.
\end{eqnarray}

By subtracting (\ref{eq:prima}) from (\ref{eq:seconda}) we obtain 
\[\epsilon(\hat{x}_{j}(1),\gamma(f(\hat{x}_{j}(1)),x_j(0)))=\epsilon(\hat{x}_{j}(1),\gamma(g(\hat{x}_{j}(1)),x_j(0))).\]
Since we can repeat the argument for any $j\in I$, the lemma follows.  
\begin{flushright}
$\square$
\end{flushright}

\section{The theorem}
\label{sec:theorem}

Let us denote by $f\vee g$ the component-wise maximum of $f$ and $g$. Then we have:
\begin{theorem} \label{thm}
Let $\tilde{f}$ and $\tilde{g}$ be second-order reversible cellular automata of neighborhood size $\beta$ and reversible combiner $\gamma$, and suppose $\epsilon=\epsilon(\hat{x}_i(t),x_i(t-1))$ is an additive conserved density of size $\alpha$ for $\tilde{f}$ and $\tilde{g}$.  Then $\epsilon$ is an additive conserved density also for $\tilde{h}$, where $h=f \vee g$.
\end{theorem}
{\it Proof.} Suppose that initial conditions 
\[x(0)=\{x_0(0),x_1(0),\dots,x_{N-1}(0)\}\]
 and 
 \[x(1)=\{x_0(1),x_1(1),\dots,x_{N-1}(1)\}\] are given.
Using the notation introduced in the previous section, we need to show that
\[\Delta'_{N,f}(x(1),x(0))=\Delta'_{N,f\vee g}(x(1),x(0))=0.\]
This amounts to showing that
\begin{equation}   \label{eq : Eq.3}
\sum_{i=0}^{N-1}\epsilon(\hat{x}_{i}(0),x_{i}(1))=\sum_{i=0}^{N-1}\epsilon(\hat{x}_i(1),\gamma((f\vee g)(\hat{x}_i(1)),x_{i}(0))).
\end{equation}

First notice that we can write the right-hand side of Eq.(\ref{eq : Eq.3}) as 
\begin{eqnarray}  \label{eq : Eq.4}
&&\sum_{i\in\Lambda_1}\epsilon(\hat{x}_i(1),\gamma(f(\hat{x}_i(1)),x_{i}(0)))
\nonumber\\
&&\phantom{aaa}+\sum_{i\in \Lambda_2}\epsilon(\hat{x}_i(1),\gamma(g(\hat{x}_i(1)),x_{i}(0)))
\nonumber\\
&&\phantom{aaaaaa}+\sum_{i\in \Lambda_3}\epsilon(\hat{x}_i(1),\gamma(f(\hat{x}_i(1)),x_{i}(0))),
\end{eqnarray}
where
\begin{eqnarray*}
\Lambda_1 &:=& \{i\ | f(\hat{x}_i(1))>g(\hat{x}_i(1)) \}\\
\Lambda_2 &:=& \{i\ | f(\hat{x}_i(1))<g(\hat{x}_i(1)) \}\\
\Lambda_3 &:=& \{i\ | f(\hat{x}_i(1))=g(\hat{x}_i(1)) \}.
\end{eqnarray*}
By Lemma~\ref{prop:prop1} we have that
\begin{equation}
\epsilon(\hat{b}(i,1),\bfgm(g(\hat{b}(i,1)),b(i,0)))=\epsilon(\hat{b}(i,1),\bfgm(f(\hat{b}(i,1)),b(i,0))),
\end{equation}
and therefore the expression in (\ref{eq : Eq.4}) is equal to:
\begin{eqnarray}\nonumber
\sum_{i\in\Lambda_1}\epsilon(\hat{b}(i,1),\bfgm(f(\hat{b}(i,1)),b(i,0)))
+\sum_{i\in \Lambda_2}\epsilon(\hat{b}(i,1),\bfgm(f(\hat{b}(i,1)),b(i,0)))+\\
\sum_{i\in \Lambda_3}\epsilon(\hat{b}(i,1),\bfgm(f(\hat{b}(i,1)),b(i,0)))=\nonumber
\sum_{i=0}^{N-1}\epsilon(\hat{b}(i,1),\bfgm(f(\hat{b}(i,1)),b(i,0))). 
\end{eqnarray}
Since $\epsilon$ is a density conserved quantity for $f$ the above is equal to
\[\sum_{i=0}^{N-1}\epsilon(\hat{b}(i,0),b(i,1)),\]
and therefore we have that 
\[\sum_{i=0}^{N-1}\epsilon(\hat{b}(i,0),b(i,1))=\sum_{i=0}^{N-1}\epsilon(\hat{b}(i,1),\bfgm((f\vee g)(\hat{b}(i,1)),b(i,0))).
\]
\begin{flushright}
$\square$
\end{flushright}

\begin{example}

As the size of the alphabet increases, the number of possible reversible combiners undergoes a combinatorial explosion, so computing the equivalence set $\mathcal{F}_I$ becomes computationally impossible.  What we can do, however, is to fix a reversible combiner and hunt for pairs $(f,g)$ of CA with the same conserved quantities and check the validity of the theorem by observing that $\psi(f)\subseteq\psi(f\vee g)$. 

For instance, for $|S|=3$ and the reversible combiner $\gamma$ such that $\gamma $(0, 0) = 0; $\gamma $(1, 0) = 1; $\gamma $(2, 0) = 2; $\gamma $(0, 1) = 1; 
$\gamma $(1, 1) = 0; $\gamma $(2, 1) = 2; $\gamma $(0, 2) = 2; $\gamma $(1, 2) = 1; $\gamma $(2, 2) = 0, we found that the CA (expressed in Wolfram's notation in base 3) $1311051521973$ and $2889590638889$ belong to the same class.  Then we computed $1311051521973 \vee 2889590638889= 6589964923167$ and verified that $\psi(1311051521973)\subseteq\psi(6589964923167)$.
\end{example}

\begin{remark}
The substance of the theorem is that of proving the existence of an order relation on CA which is preserved by  the conserved quantities.  Consider the set $\mathcal{F}_I$ of second-order reversible CA with a fixed window $I$ with the partial order relation induced by $\vee$: 
 \[f< g\iff f\vee g=g.\]
 Then, if we consider the set $\langle C\rangle=\{f\vee g\}_{f,g\in C}$ generated by any equivalence class $C\in \mathcal{F}_I/\sim$, the theorem says that $\psi$ is an order-preserving map. Indeed, since there is a natural partial order $\subseteq$ between subspaces of $\mathbb{R}^n$, we have that 
\[f<g\Rightarrow \psi(f)\subseteq \psi(g).\]
\end{remark}

\section{Conclusions}
In this article we have studied a certain class of second-order, reversible CA that were first considered by Takesue \cite{Takesue} for their ability to exhibit thermodynamic behavior. 

The inspiration for this work has, in part, come from Noether's theorem, which is one of the most celebrated results both in mathematics and  physics. It asserts that conserved quantities are in correspondence with symmetries of the laws of nature. Time-translation symmetry is associated with conservation of energy, space-translation symmetry with conservation of momentum, and rotation-symmetry is associated with conservation of angular momentum. 

These results hold under the assumption that space and time are both continuous: to what extent do they still hold for discrete systems such as CA? The answer to this question is still an open problem, and a fully satisfactory discrete version of Noether's theorem has not yet been formulated. In this work, however, we show that there exists a logical organization of additive conserved quantities which reflects some symmetries of the system, at least for a large class of second-order, reversible CA. In particular, we have shown that there is a partial order relation defined on CA which is preserved by the set of their respective conserved quantities. 

There are still many open problems in the theory of reversible CA which we hope to approach using our result. In particular, there is an interesting conjecture claimed in \cite{hattak} which states that  ``antisymmetric conserved quantities'', that is those such that $E(x(t), x(t+1))=-E(x(t+1), x(t))$, can only be associated to CA endowed with special symmetries (Rule 90, for instance, is one of them). More generally, it appears that, even in the discrete case, it is possible to classify conserved quantities according to symmetry groups.

As in a previous paper of ours \cite{BoghosianCaterina}, we have herein suggested an approach to this problem which only uses combinatorial techniques.  We hope that this work will contribute towards the formulation of a discrete version of Noether's theorem, which so neatly classifies the conserved quantities of continuous systems according to the characteristics of their dynamics.

\section*{Acknowledgments}

This work was partially funded by ARO award number W911NF-04-1-0334, AFOSR award number FA9550410176, and facilitated by scientific visualization equipment funded by NSF award number 0619447.  The authors are grateful to Peter Love and Zbigniew Nitecki for helpful conversations.

\end{document}